\documentclass{article}
\usepackage{frascatiphys}
\newcommand{\open}{{<\kern -0.3 em{\scriptscriptstyle )}}}
\begin{document}
\title{ 
HOW TO STUDY QUARK SPIN WITHOUT SPIN
}
\author{
Marco Radici        \\
{\em Istituto Nazionale di Fisica Nucleare - Sezione di Pavia - Pavia - Italy} \\
}
\maketitle
\baselineskip=11.6pt
\begin{abstract}
I review the most important single- and double-spin asymmetries that allow for
the extraction of transversity and other chiral-odd and/or T-odd parton
densities, necessary to explore the partonic content and the spin structure of
the nucleon. With particular reference to the proposed GSI-HESR facility, I
report on some Monte-Carlo simulations of cross sections and spin asymmetries for
(un)polarized Drell-Yan with protons and antiprotons at the proposed kinematics
for this future facility.
\end{abstract}
\baselineskip=14pt
\section{Introduction}
\label{sec:intro}
Recently, several experimental collaborations reported on nonvanishing
single-spin asymmetries (SSA) in hard processes involving a target with a
nonnegligible percentage of transverse polarization 
${\bf S}_{_T}$~\cite{smc,hermes}. Historically, the first discussion about 
transverse-spin effects in high-energy physics goes back to the late seventies, 
when an anomalous large transverse polarization of the $\Lambda$ produced in $pN$ 
annihilations was measured, surviving even at large values of transverse momentum 
$p_{_T}$~\cite{fermilab}. Such an observation requires a nonvanishing imaginary 
part in the off-diagonal part of the fragmentation matrix of quarks into 
$\Lambda$, which is forbidden in QCD at leading twist and appears as a 
${\cal O}(1/p_{_T})$ effect~\cite{pqcd}. A pioneering work soon 
appeared~\cite{poldy} about the possibility of having leading-twist asymmetries in 
fully polarized Drell-Yan processes, but it was basically ignored for almost a 
decade upon the prejudice that transverse-spin effects have to be suppressed. 

The above quoted recent observations warn us about two fundamental issues. First
of all, even the leading-order (transverse) spin structure of the nucleon (and of
hadrons, in general) is far from being fully understood. Secondly, SSA signal the
appearance of effects that cannot be explained by perturbative QCD, because they
are essentially related to a correlation between the polarization and the
nonperturbative intrinsic transverse momentum of quarks, as well as to their
orbital motion inside the parent hadron. Such features can coexist in objects
linked to processes flipping the quark helicity and/or describing residual
interactions between the quark and the surrounding hadronic matter, such that the
invariance upon time-reversal does not put any constraint. In common jargon, the
first group is made of chiral-odd functions, since at leading twist chirality and
helicity are identical\cite{jaffe-erice}; QCD does preserve helicity, so these 
functions pertain the "soft" domain where the chiral symmetry of QCD is 
(spontaneously) broken. The most important one is a parton density related to the 
distribution of the transverse polarization of quarks in transversely polarized 
hadrons. It is a leading-twist parton distribution function (PDF), therefore 
necessary to complete the knowledge of the partonic spin structure of hadrons, but 
it escaped notice so far because of its chiral-odd nature, which prevents it from 
being extracted in most common processes like inclusive Deep-Inelasitc Scattering 
(DIS)~\cite{jaffe-erice}. In Sec.~\ref{sec:h1}, I will briefly review its main 
properties (for a more thorough review see ref.~\cite{barone}). 

The second group of functions is named, in jargon, T-odd: it does not mean a
violation of the fundamental law of nature, it simply indicates that the above
mentioned residual interactions prevent  the time-reversal operation to put any
constraint on such functions. Nevertheless, for several years a common prejudice
prevented people from recognizing the existence of such objects both as PDF and
parton fragmentation functions (PFF), and led them to reject the possibility of
asymmetries in processes like $pp^\uparrow \rightarrow \pi X$ because of the
violation of invariance under time-reversal transformations. The above mentioned
recent observation of such an asymmetry clearly contradicts such prejudice. The
most famous example of T-odd functions is the Collins function~\cite{collins}, 
that I will recall in Sec.~\ref{sec:Collins} together with other similar objects, 
the interference fragmentation functions (IFF)~\cite{jaffe-iff,noi1}.

The rich scenario depicted above, obtained by releasing some of the contraints
introduced by standard perturbative QCD, lead to even more "exotic", but useful,
conclusions. In fact, the existence of chiral-odd and/or T-odd PDF (PFF) allows
for the description of the polarization state of quarks irrespectively of the
polarization state of the parent hadron; it is then truly possible to perform
spin physics without using polarized targets. In Sec.~\ref{sec:dy}, I will recall
how the combination of unpolarized and polarized Drell-Yan processes,
particularly involving antiprotons, can help us in disentangling the unknown PDF
and PFF occurring elsewhere, like in semi-inclusive DIS (SIDIS) or proton
annihilation. In the last section, Sec.~\ref{sec:mc}, I will explicitly show some
Monte-Carlo simulations for such Drell-Yan processes in the kinematics of
interest for the GSI-HESR future facility.

\section{Transversity}
\label{sec:h1}
\begin{figure}[h]
 \vspace{3.0cm}
\includegraphics{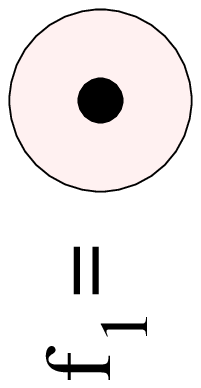} 
\hspace{1truecm} 
\includegraphics{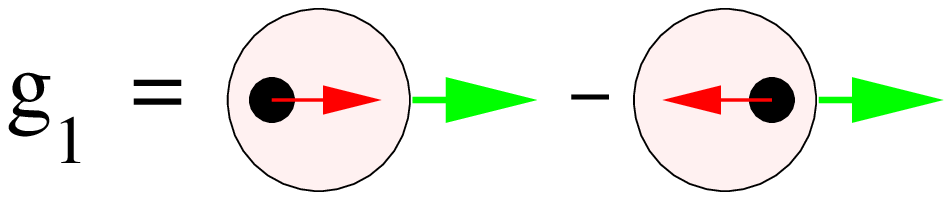} \\
\begin{center}
\includegraphics{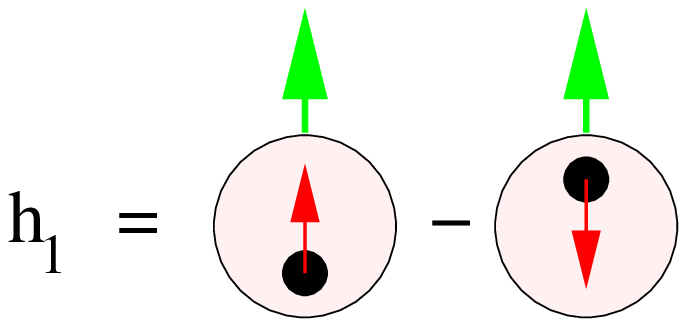}
\end{center}
\caption{\it The probabilistic interpretation of $f_1, g_1,$ and $h_1$,
respectively.
\label{fig:f1g1h1} }
\end{figure}

At leading twist, three PDF are needed to describe the partonic spin structure of
the nucleon. The easiest and most intuitive way to see it is to expand the PDF on
the quark-nucleon helicity basis~\cite{jaffe-iff}, 
\begin{eqnarray}
{\rm PDF}(x,Q^2) &= &\frac{1}{2}\,q(x,Q^2)\,I \otimes I + \frac{1}{2}\, \Delta 
q(x,Q^2) \, \sigma_3 \otimes \sigma_3 \nonumber \\
& &\quad + \frac{1}{2}\,\delta q(x,Q^2)\,\left[ \sigma_+ \otimes \sigma_- + 
\sigma_- \otimes \sigma_+ \right] \; . 
\label{eq:pdf-hel}
\end{eqnarray}
\begin{figure}[t]
 \vspace{5.0cm}
\includegraphics{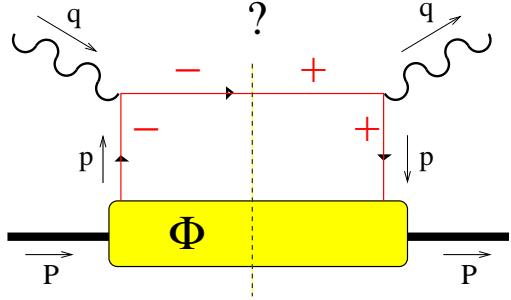}
\caption{\it The main contribution to inclusive DIS. Intermediate quark helicity
states are also indicated. 
\label{fig:dis} }
\end{figure}
The functions $q,\Delta q,$ are the momentum and helicity distributions, which
are well known experimentally and have a clear probabilistic interpretation. The
function $\delta q$ is not diagonal in the helicity basis; it mixes different
helicity (hence, chirality) states and, therefore, it is suppressed in one of the
simplest measurable processes, the inclusive DIS (fig.~\ref{fig:dis}). For this
reason it escaped notice until recently. But it is a fundamental piece of
infomation about the nucleon spin structure, with the same dignity as the other
two PDF. In fact, by changing from the helicity to the transverse spin basis, the
role of $\Delta q$ and $\delta q$ is interchanged; now, $\delta q$ is diagonal
and can be interpreted as the probability to find a quark with its spin polarized
along the transverse spin of a polarized nucleon minus the probability to find it
polarized oppositely~\cite{jaffe-erice}. In the following, I will switch to the 
more common notations of $f_1,g_1$ and $h_1$ for the three PDF discussed above: 
$f$ will always indicate unpolarized partons, while by $g,h,$ I mean their 
longitudinal and transverse polarization, respectively; finally, the index $_1$ 
indicates that these PDF happen at leading twist. The situation is graphically 
summarized in fig.~\ref{fig:f1g1h1}.

Contrary to the other two PDF, the transversity $h_1$ does not have a
counterpart in the Quark Parton Model (QPM) at the level of DIS structure
function, because of its chiral-odd nature. Even if it depends on spin, it is not
related to a partonic fraction of the nucleon spin, because the related twist-2
operator is not part of the full angular momentum tensor. The first moment of
$h_1^f$, for a quark with flavor $f$ in a nucleon state $|PS\rangle$ with
momentum $P$ and spin $S$, is called the tensor charge, i.e.
\begin{eqnarray}
\langle PS|\bar{q}^f i \sigma^{0i}\gamma_5 q^f|PS\rangle\Bigg\vert_{Q^2} &= &
2 S^i \int dx \left[ h_1^f(x,Q^2) - \bar{h}_1^f(x,Q^2)\right] \nonumber \\
&= &2S^i \delta q^f (Q^2) \, \sim \, \frac{1}{(\log Q^2)^\gamma} \; ,
\label{eq:tens-ch}
\end{eqnarray}
because it is related to the tensor operator $\sigma^{\mu\nu} \gamma_5$; it has a
nonvanishing anomalous dimension $\gamma$ and, therefore, from a renormalization
scale $Q^2$ it evolves unavoidably to zero~\cite{jaffe-erice}. On the contrary, 
the nonsinglet axial charge $2S^i \Delta q^f (Q^2)$ is directly related to the
nucleon axial charge $g_{_A}$ so that the corresponding distribution $g_1(x,Q^2)$
truly describes the quark helicity as a fraction of the nucleon's one. The other
big difference can be read from eq.~\ref{eq:tens-ch}, namely the transversity for
the antiquark enters with an opposite sign such that $\delta q^f$ is odd under
charge conjugation operations, contrary to $\Delta q^f$. This means that, under
evolution, $h_1$ does not mix with charge-even structures like the $q\bar{q}$
pairs from the Dirac sea. If we now realize that $\delta q^f$ cannot mix also
with gluonic operators, which are both chiral- and charge-even in a spin
${\textstyle \frac{1}{2}}$ hadron, we come to the conclusion that the
transversity has very peculiar evolution properties: each moment does depend on
the scale $Q^2$, but the function evolves homogeneously with $Q^2$ like a pure
nonsinglet structure function, completely decoupled from gluons and $q\bar{q}$
pairs~\cite{jaffe-erice}. It is probably the best place to explore the valence 
quark content of the nucleon and to test models based on the concept of 
constituent quark. 

From eq.~\ref{eq:pdf-hel} it is evident that $g_1$ is associated to the operator
$\sigma_3$ while $h_1$ to $\sigma_1$ (or $\sigma_2$ by a rotation in the spin
space around $\hat z$). Since in the nonrelativistic framework spin and space
operations (Euclidean boosts, etc..) commute, we easily get $g_1=f_1$, as it is
also graphically intuitive from fig.~\ref{fig:f1g1h1}. Therefore, any deviation
from this equality will tell us about the relativistic nature of quark motion in
the nucleons, as much as the nonrelativistic prediction $g_{_A} = {\textstyle
\frac{5}{3}}$ deviates from the experimental value of 
$1.255 \pm 0.006$~\cite{pdg}. 

Finally, from the positivity of probability distributions we get
\begin{eqnarray}
|g_1^f(x,Q^2)| &\leq &f_1^f (x,Q^2) \nonumber \\
|h_1^f(x,Q^2)| &\leq &f_1^f (x,Q^2) \; ,
\label{eq:diseq}
\end{eqnarray}
while from the quark distribution matrix of eq.~\ref{eq:pdf-hel} in helicity
basis being semi-positive definite, we get the socalled Soffer
inequality~\cite{soffer-ineq}, 
\begin{equation}
f_1^f(x,Q^2) + g_1^f(x,Q^2) \geq 2|h_1^f(x,Q^2)| \; ,
\label{eq:soffer}
\end{equation}
which holds up to next-to-leading order in QCD radiative
corrections~\cite{soffer-ineqNLO,soffer-ineqNLO2}.

\section{Transversity from semi-inclusive processes}
\label{sec:Collins}
Since the transversity $h_1$ is a chiral-odd PDF happening at leading twist, we
need to find a leading-twist chiral-odd partner in order to extract it from the
cross section. There are two possibilities: a polarized Drell-Yan process where
$h_1$ happens in combination with the transversity $\bar{h}_1$ for an antiquark, 
or a semi-inclusive process, where $h_1$ is combined with a chiral-odd PFF. 
Despite the fact that the extraction of $h_1$ from Drell-Yan was suggested as 
first in the literature long time ago~\cite{poldy}, we will postpone it to the 
next sec.~\ref{sec:dy}. Here in the following, we will discuss SIDIS or 
annihilation processes leading to the Collins effect and other interesting 
phenomena.

\begin{figure}[t]
 \vspace{5.0cm}
\includegraphics{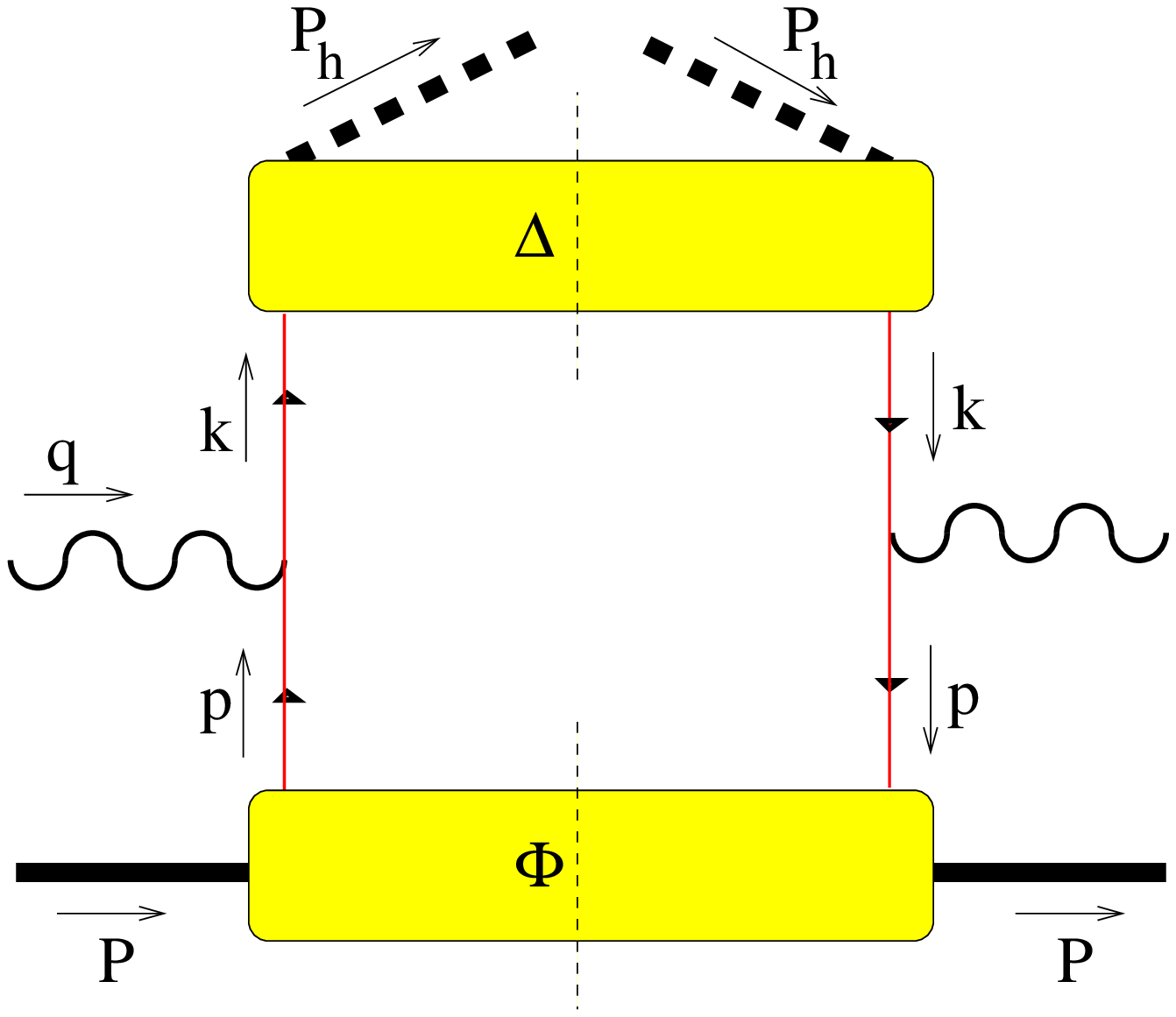} 
\hspace{1truecm} 
\includegraphics{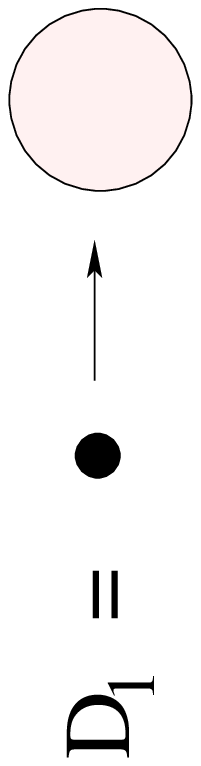} \\
\includegraphics{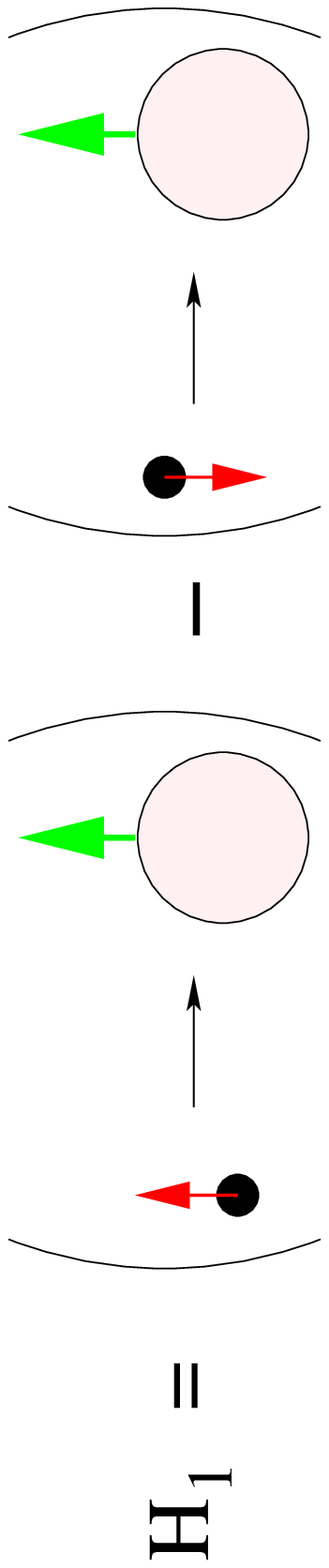}
\caption{\it The handbag diagram, the leading contribution to SIDIS, and the
probabilistic interpretation of $D_1$ and $H_1$.
\label{fig:D1H1} }
\end{figure}

\subsection{One-hadron semi-inclusive processes}
\label{sec:sidis}
Since we are looking for a semi-inclusive process where transverse polarization is
involved at the partonic level, we are naturally led to select a final state with a
transversely polarized hadron. The most natural choice is the hyperon $\Lambda$,
whose polarization is easily deduced from the angular distribution of its decay
products. For the electroproduction of transversely polarized $\Lambda$ on a
transversely polarized proton, $ep^\uparrow \rightarrow e'\Lambda^\uparrow X$, the
leading contribution comes from the well known handbag diagram (see
fig.~\ref{fig:D1H1}); $h_1$ can then be extracted by the double-spin asymmetry 
(DSA, or depolarization or spin transfer coefficient, as differently reported in 
the literature)~\cite{jaffe-lambda}
\begin{equation}
D_{NN} = \frac{d\sigma (p^\uparrow \Lambda^\uparrow) - d\sigma (p^\downarrow
\Lambda^\uparrow)}{d\sigma (p^\uparrow \Lambda^\uparrow) + d\sigma (p^\downarrow
\Lambda^\uparrow)} \, \propto \, |{\bf S}_{_T}||{\bf S}_{\Lambda_T}|\, 
\frac{\sum_f\,e_f^2\,h_1^f(x)\,H_1^f(z)}{\sum_f\,e_f^2\,f_1^f(x)\,D_1^f(z)} \; ,
\label{eq:dsa}
\end{equation}
where $D_1$ is the PFF of an unpolarized quark into the unpolarized $\Lambda$ and
$H_1$ is its partner for the transversely polarized case (see fig.~\ref{fig:D1H1};
the notations follow the same convention as for the PDF, but with upper case 
letters and with the exception of $D_1$, in order not to be confused with the DIS 
structure function $F_1$). On the experimental side, the DSA has been measured 
also for the reaction $pp^\uparrow \rightarrow \Lambda^\uparrow X$~\cite{dsa}, 
which has the same partonic content as eq.~\ref{eq:dsa} but for an additional 
$\bar{f}_1$ for the annihilating unpolarized antiquark. On the theoretical side, 
the problem is related to the identification of the spin transfer mechanism: while 
it is somewhat clear how to reproduce the $\Lambda$ properties from the $uds$ 
valence picture using SU$_f(3)$, it is not at all clear how the transverse 
polarization of the fragmenting quark is transferred to the transversely polarized 
$\Lambda$. Several models are available in the literature (for example, see 
ref.~\cite{to-lambda}); the use of polarized proton and antiproton beams at GSI 
would help in selecting them, also because the PDF of an antiquark is not 
suppressed in an antiproton. 

\begin{figure}[t]
 \vspace{4.0cm}
\includegraphics{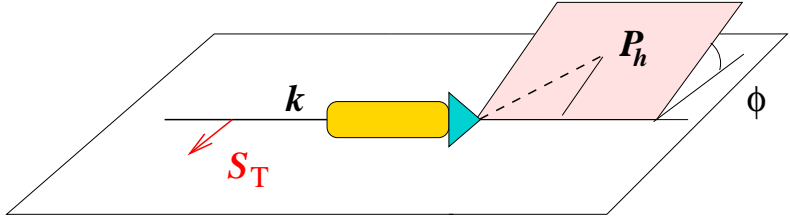} 
\caption{\it The Collins mechanism.
\label{fig:collins} }
\end{figure}

\begin{figure}[t]
 \vspace{3.0cm}
\includegraphics{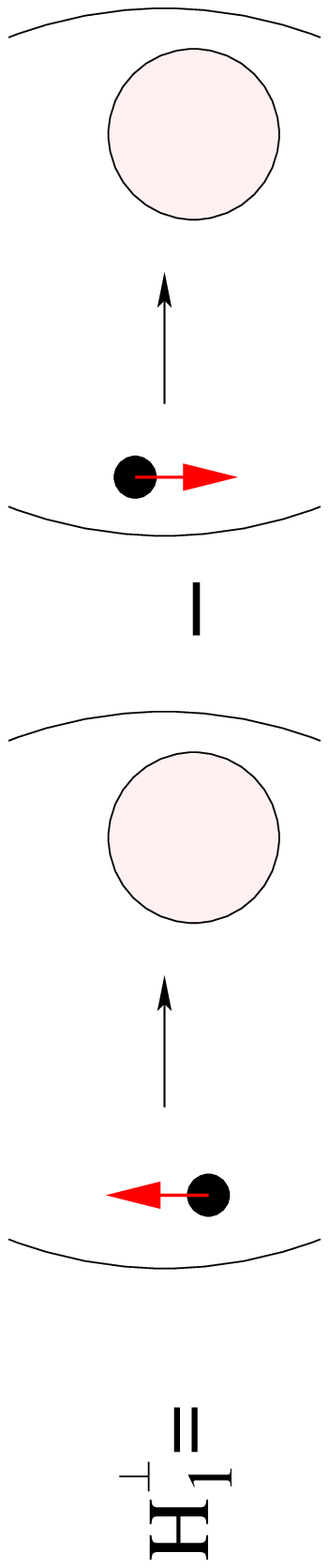} 
\caption{\it The Collins function.
\label{fig:H1perp} }
\end{figure}

The second option is to transfer the transverse polarization of a fragmenting quark
to an unpolarized hadron with an explicit dependence on its transverse momentum
${\bf P}_{h\perp}$. The leading contribution is again given by the handbag diagram
of fig.~\ref{fig:D1H1}, but the azimuthal asymmetry is now determined by the T-odd
mixed product $\sin \phi_{_C} \propto {\bf k}\times {\bf P}_{h\perp} \cdot {\bf
S}_{_T}$~\cite{collins}, as represented in fig.~\ref{fig:collins}, where 
$\phi_{_C}$ is the socalled Collins angle. An explicit dependence of the cross 
section on ${\bf P}_{h\perp}$ implies a sensitivity to the transverse momenta of 
the partons involved in the hard vertex. For the handbag diagram of 
fig.~\ref{fig:D1H1}, the leading-twist decomposition of the quark-quark 
correlators $\Phi$ and $\Delta$ depending explicitly on the quark transverse 
momenta ${\bf p}_{_T}$ and ${\bf k}_{_T}$, respectively, shows a very rich 
structure~\cite{bible}, where $h_1(x,{\bf p}_{_T})$ appears convoluted with the 
Collins function $H_1^\perp (z,{\bf k}_{_T})$, a chiral-odd and T-odd PFF whose 
probabilistic interpretation is depicted in fig.~\ref{fig:H1perp} (the "$\perp$" 
notation indicates that the Collins function appears weighted by the transverse 
momentum ${\bf k}_{_T}$). 

Such a SSA has been observed at SMC for the $pp^\uparrow \rightarrow \pi X$
reaction, but with a very low resolution in the final state~\cite{smc}. The 
Collins effect has been observed also for the process $ep^\uparrow\rightarrow 
e'\pi X$~\cite{hermes}, but using a target polarized along the beam direction 
with a consequent small fraction of transverse polarization with respect to the 
momentum transfer $q$. For this setup, both the cross section for longitudinally 
and transversely polarized targets do contribute at leading and subleading twist, 
since $|{\bf S}_{_T}|$ is suppressed by a factor $1/Q$ with respect to the 
helicity $\lambda$. Several contributions mix up in the SSA and it is not easy to 
select the combination corresponding to the Collins effect (for a detailed 
discussion, see for example ref.~\cite{oga}).

Only recently, data were taken at HERMES with a pure transversely polarized 
target. In this case, the six-fold differential cross section reads~\cite{bible2}
\begin{eqnarray}
\frac{d^6\sigma_{_{OT}}}{dx dy dz d\phi_{_S}d{\bf P}_{h\perp}} &\propto &|{\bf
S}_{_T}|\, \Bigg \{ \sin (\phi_h + \phi_{_S}) \, {\cal F}\, \left[ 
\frac{{\bf k}_{_T}\cdot {\hat h}}{M_h}\, x h_1(x,{\bf p}_{_T}^2)\, H_1^\perp (z,
{\bf k}_{_T}^2)\right] \nonumber \\
& &+ \sin (\phi_h - \phi_{_S}) \, {\cal F}\, \left[ \frac{{\bf p}_{_T}\cdot 
{\hat h}}{M}\,x f_{1T}^\perp (x,{\bf p}_{_T}^2)\, D_1^\perp (z, {\bf k}_{_T}^2)
\right] \nonumber \\
& &+ \sin (3\phi_h - \phi_{_S}) \, {\cal F}\, \left[ a({\bf p}_{_T}, {\bf k}_{_T})
\, h_{1T}^\perp (x,{\bf p}_{_T}^2)\,  H_1^\perp (z,{\bf k}_{_T}^2)\right] \; ,
\nonumber \\
& &\label{eq:ssa}
\end{eqnarray}
where $M,M_h$ are the proton and pion masses, respectively, and the convolution is
defined as
\begin{equation}
{\cal F} [...] \equiv \int d{\bf p}_{_T} d{\bf k}_{_T}\, \delta ({\bf p}_{_T} + 
{\bf q}_{_T} - {\bf k}_{_T}) ... \; , 
\label{eq:convol}
\end{equation}
with ${\bf q}_{_T}$ the transverse component of the momentum transfer in the hard
vertex. If the lab plane is defined by the directions of the beam and of the target
polarization vector, then the azimuthal angles $\phi_s, \phi_h$ in 
eq.~\ref{eq:ssa} represent the orientation of the lab plane and final hadronic 
plane (formed by ${\bf P}_{h\perp}$ and ${\bf q}$) with respect to the scattering 
plane, respectively. The contribution corresponding to the Collins effect 
($\sin \phi_{_C} \equiv \sin (\phi_s + \phi_h)$) can be extracted by the 
following SSA~\cite{bible2}
\begin{eqnarray}
\langle \frac{{\bf P}_{h\perp}}{M_h}\,\sin \phi_{_C}\rangle &\equiv &
\frac{\int d\phi_s d{\bf P}_{h\perp}\, \sin \phi_{_C}\, [d\sigma(p^\uparrow) -
d\sigma(p^\downarrow)]}{\int d\phi_s d{\bf P}_{h\perp}\, [d\sigma(p^\uparrow) +
d\sigma(p^\downarrow)]} \nonumber \\
&\propto &|{\bf S}_{_T}|\,\frac{\sum_f\,e_f^2\,xh_1^f(x)\,H_1^{\perp
f(1)}(z)}{\sum_f \, e_f^2\, f_1^f(x)\, D_1^f(z)} \; , 
\label{eq:wssa}
\end{eqnarray}
where $H_1^{\perp f(1)}(z) = \int d{\bf k}_{_T}\, \frac{{\bf k}_{_T}^2}{2M_h}\, 
H_1^\perp (z,{\bf k}_{_T}^2)$ is the first moment of the Collins function (the
simple $\langle \sin \phi_{_C} \rangle$ does not break the convolution, unless some
assumption on the ${\bf p}_{_T}$ and ${\bf k}_{_T}$ dependence of the functions is
made). Eq.~\ref{eq:wssa} implies that, when calculating the radiative corrections in
order to determine its QCD evolution, the explicit ${\bf P}_{h\perp}$ dependence
breaks the collinear factorization and forces to resum all the contributions of real
and virtual soft gluons into the socalled Sudakov form factors, which can largely
dilute the SSA~\cite{sudakov}. The same problem appears when trying to extract 
information on the Collins function via the corresponding $e^+e^- \rightarrow \pi^+ 
\pi^- X$ reaction, as it is somewhat confirmed by the analysis of the DELPHI data 
collection where an asymmetry of at most 6\% is observed~\cite{delphi} (but with 
large uncertainties). 

\begin{figure}[t]
\vspace{3.0cm}
\includegraphics{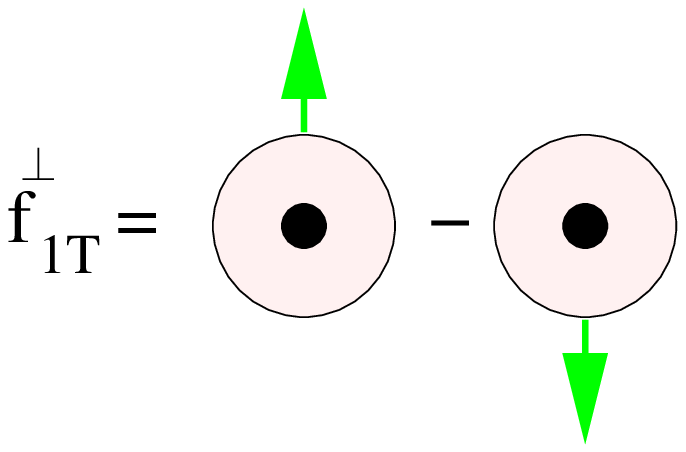} 
\hspace{0.5cm}
\includegraphics{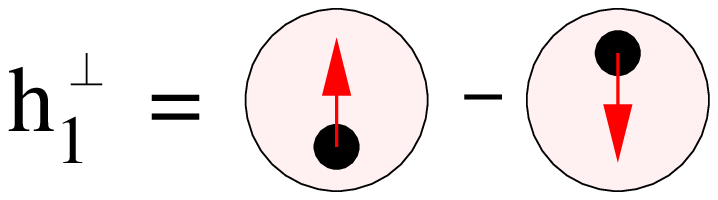} 
\caption{\it The Sivers and the "Boer" functions on the left and the right,
respectively.
\label{fig:f1Tperph1perp} }
\end{figure}

The second term in eq.~\ref{eq:ssa} represents the socalled Sivers
effect~\cite{sivers}, because it is driven by the chiral-even T-odd Sivers function 
$f_{_{1T}}^\perp$ that describes how the distribution of unpolarized quarks is 
affected by the transverse polarization of the parent hadron (see 
fig.~\ref{fig:f1Tperph1perp}), from which the notation "$_{_T}$" originates. As it is 
evident from its azimuthal dependence, this effect can be disentangled by the 
Collins one only if the target polarization vector sticks out of the scattering 
plane, i.e. $\phi_s \neq 0$. The situation is potentially more confused in the 
annihilation process $pp^\uparrow \rightarrow \pi X$, where the Collins and Sivers 
effects do not exhaust all the possibilities. In fact, at leading twist one more 
combination generates a similar azimuthal asymmetry, namely the SSA~\cite{boer}
\begin{equation}
\frac{d\sigma(p^\uparrow)-d\sigma(p^\downarrow)}{d\sigma(p^\uparrow)+
d\sigma(p^\downarrow)} \, \propto \, \frac{\sum_f\,h_1^{\perp f}(x_1)\, \bar{h}_1^f
(x_2)\, D_1^f(z)}{\sum_f \, f_1^f(x_1)\,\bar{f}_1^f(x_2)\,D_1^f(z)} \; ,
\label{eq:ssa-ann}
\end{equation}
where $h_1^\perp$ is a twist-2 chiral-odd T-odd PDF describing the influence of the
quark transverse polarization on its momentum distribution inside an unpolarized
parent hadron (see fig.~\ref{fig:f1Tperph1perp}). Moreover, even if the involved PDF
appear at leading twist, the required explicit dependence of the elementary cross
section on transverse momenta of the partons introduces a suppression factor,
raising the relative importance of subleading-twist effects like the Qiu-Stermann
one~\cite{qiu}.

\subsection{Color-gauge invariance and T-odd parton densities}
\label{sec:color}
In the previous section, we encountered several examples of T-odd PDF and PFF that
appear as soon as we allow them to depend explicitly on the parton transverse
momentum; in fact, integration upon the latter washes all these effects away. 

We already mentioned that the jargon "T-odd" indicates no constraints on the
considered function from the invariance under time-reversal transformation. For
the Collins function, this can be simply interpreted by advocating residual Final
State Interactions (FSI) occurring between the detected hadron and the residual
jet. As soon as we reduce the hadron wave function to a plane wave, the
time-reversal invariance forbids the Collins effect and makes $H_1^\perp$ to
vanish; but in reality there are FSI between this hadron and the jet on a time
scale much longer than the hard vertex, hence the $in$ and $out$ states cannot be
interchanged and the invariance under time-reversal does not put any constraint:
T-odd structures are allowed and we observe the Collins effect. 

\begin{figure}[t]
 \vspace{4.0cm}
\includegraphics{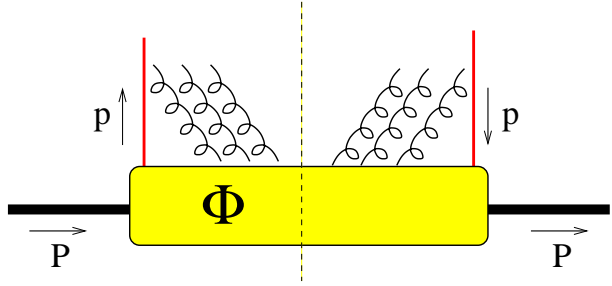}
\caption{\it A possible representation of the color-gauge invariant quark-quark
correlator.
\label{fig:link} }
\end{figure}

But how to interpret the T-odd PDF $f_{_{1T}}^\perp$ and $h_1^\perp$? If for the
$pp^\uparrow \rightarrow \pi X$ process we can think about a sort of Initial
State Interaction (ISI) occurring before the hard 
annihilation~\cite{brodsky,collins2,jiyuan}, we cannot recycle the same idea in the 
SIDIS process $ep^\uparrow \rightarrow e'\pi X$. However, there is another 
"technical" consideration that leads to T-odd structures. The PDF can be obtained 
as suitable Dirac projections of the quark-quark correlator
\begin{equation}
\Phi (x,S) = \int \frac{d\zeta^-}{2\pi}\,e^{ixP^+\zeta^-}\,\langle P,S|
\bar{\psi}(0)\, U_{[0,\zeta^-]}\, \psi(\zeta)|P,S\rangle \Bigg\vert_{\zeta^+ =
\vec \zeta_{_T} = 0} \; , 
\label{eq:phi}
\end{equation}
which is made color-gauge invariant by the introduction of the socalled gauge
link operator
\begin{equation}
U_{[0,\zeta]} = {\cal P}\, e^{-ig \int_0^\zeta dw \cdot A(w)} \; ,
\label{eq:link}
\end{equation}
linking the two different space-time points $0,\zeta,$ by all the possible
ordered paths (${\cal P}[..]$) followed by the gluon field $A$, which couples to
the quark field $\psi$ by the constant $g$. Since the interaction in
eq.~\ref{eq:phi} runs along the suppressed "-" direction, the longitudinal $A^+$
component of the gluon field appears at any power in the expansion of the
exponential; therefore, a possible representation of the gauge link operator is
depicted in the diagram of fig.~\ref{fig:link}. Hence, a sort of residual FSI occurs
between the parton and the residual hadron, opening the possibility for T-odd
structures~\cite{link-amst}. When also the ${\bf p}_{_T}$ dependence is explicitly 
kept in eq.~\ref{eq:phi}, the path required to connect at leading twist the points
$[0,\zeta]$ is more complicated and involves also the ${\bf A}_{_T}$ component of
the gluon field~\cite{link-amst}, but the bulk of the message is unchanged. 

\subsection{Interference Fragmentation Functions}
\label{sec:iff}
\begin{figure}[t]
 \vspace{4.0cm}
\includegraphics{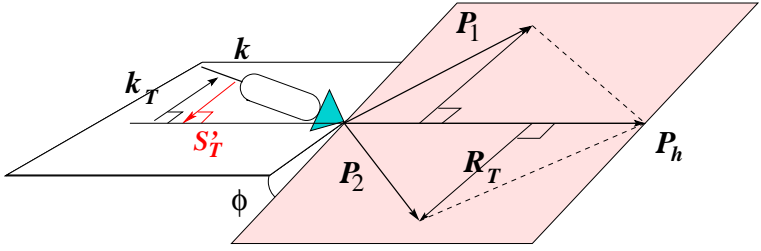}
\caption{\it The Collins-Ladinski (CL) mechanism.
\label{fig:noi} }
\end{figure}

\begin{figure}[t]
 \vspace{5.0cm}
\includegraphics{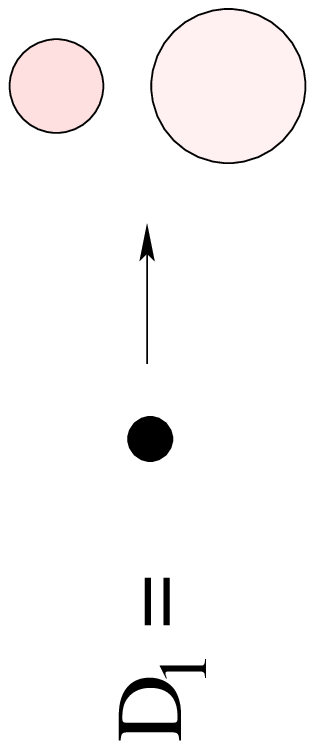} 
\hspace{1truecm}
\includegraphics{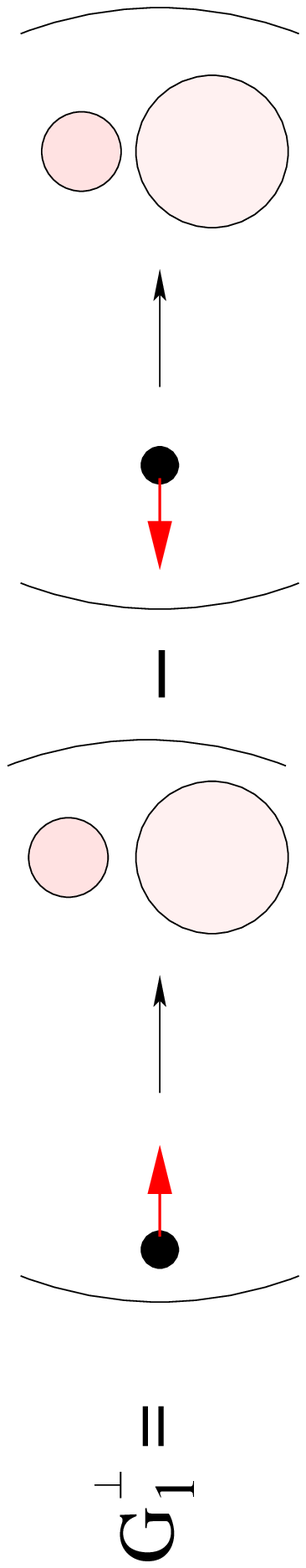} \\
\includegraphics{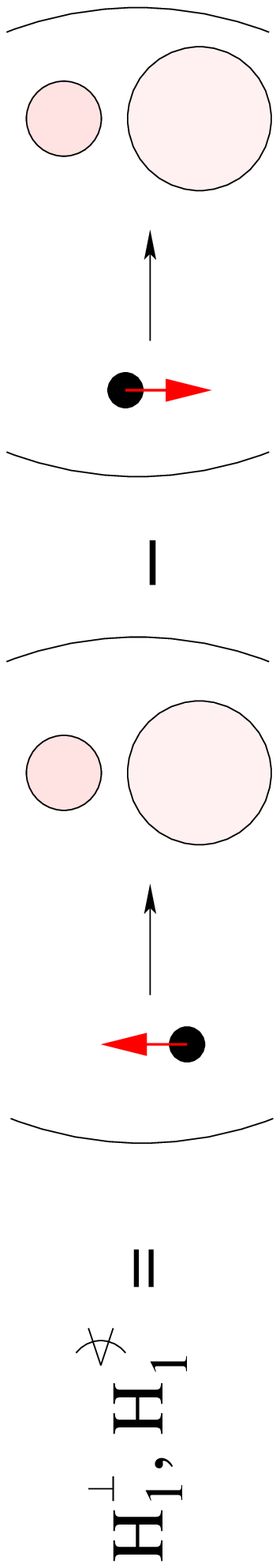}	 
\caption{\it The Interference Fragmentation Functions at leading twist.
\label{fig:iff} }
\end{figure}

In sec.~\ref{sec:sidis}, we stressed the difficulties arising when the cross
section must be kept differential also on the transverse momentum of the
detected hadron, namely the appearance of other effects, contributing to the same
asymmetry as the Collins effect, or the breaking of collinear factorization when
computing radiative corrections beyond leading order. For these reasons, it is
desirable to find a mechanism that leads to T-odd structures surviving the
integration upon transverse momenta. This requirement is fulfilled by the
semi-inclusive process where two leading hadrons are detected in the same jet.
The asymmetry happens, then, in the azimuthal angle $\sin \phi \propto {\bf P}_1
\times {\bf P}_2 \cdot {\bf S}_{_T} = {\bf P}_h \times {\bf R} \cdot {\bf
S}_{_T}$, where $P_h = P_1+P_2$ and $R=(P_1-P_2)/2$ are the total and relative
momentum of the final pair (see fig.~\ref{fig:noi})~\cite{coll-lad}. If the two 
hadrons are unpolarized, four PFF appear by projecting the leading-twist 
contributions of the proper quark-quark correlator $\Delta$~\cite{noi1}: the 
probability of an unpolarized quark to fragment into two unpolarized hadrons, $D_1$; 
the probability of a quark with positive helicity to fragment into two hadrons minus
the same probability but with negative helicity, $G_1^\perp$; and, finally, the
same probability difference for a transversely polarized quark, occurring in the
combination ${\bf k}_{_T} H_1^\perp + {\bf R}_{_T} H_1^{\open}$ (see
fig.~\ref{fig:iff}). More complicated structures appear at subleading twist,
involving also the quark-gluon-quark correlator~\cite{noi2}. As for the projection 
by the tensor operator, $H_1^\perp$ is the analogue of the Collins function, while
$H_1^{\open}$ is related to a truly new effect. In fact, the transverse
polarization of the fragmenting quark is transformed into the orbital relative
motion of the hadron pair via the vector ${\bf R}_{_T}$: the azimuthal
distribution of the two hadrons in the detector depends on the transverse
polarization of the quark. All these functions, which are T-odd but for $D_1$,
are related to the residual interactions between the two hadrons inside the jet;
therefore, they are usually referred to as Interference Fragmentation Functions
(IFF)~\cite{jaffe-iff,noi1}. Moreover, $H_1^\perp$ and $H_1^{\open}$ are chiral-odd. 
They have a rather complicated structure. Their functional dependence can
conveniently be chosen as the light-cone fraction of the quark momentum carried
by the hadron pair, $z = P_h^-/k^- = (P_1^- + P_2^-)/k^- = z\xi + z(1-\xi)$, the
subfraction in which this momentum is further shared inside the pair, $\xi$, and
the "geometry" of the pair in the momentum space: the "opening" of the pair
momenta, ${\bf R}_{_T}^2$, the relative position of the jet axis and the hadron
pair axis, ${\bf k}_{_T}^2$, and the relative position of the hadron pair plane
and the plane formed by the jet axis and the hadron pair axis, ${\bf
k}_{_T}\cdot {\bf R}_{_T}$~\cite{noi1} (see fig.~\ref{fig:noi}).

After integrating on ${\bf P}_{h\perp}$, the leading-twist nine-fold
differential cross section for the $ep^\uparrow \rightarrow e'(\pi \pi)X$
process becomes~\cite{noi3}
\begin{equation}
\frac{d\sigma_{_{OT}}}{dx\,dy\,dz\,d\xi\,dM_h^2\,d\phi_{_S}\,d\phi_{_R}} \, 
\propto \, \frac{|{\bf S}_{_T}||{\bf R}_{_T}|}{M_h}\, \sin (\phi_{_S} +
\phi_{_R})\, h_1(x)\,H_1^{\open}(z,\xi,M_h^2) \; , 
\label{eq:2hxsec}
\end{equation}
where the role of $\phi_h$ in eq.~\ref{eq:ssa} is here taken over by
$\phi_{_R}$, and the dependence on $M_h^2$ is due to the relation ${\bf
R}_{_T}^2 = \xi (1-\xi) M_h^2 - (1-\xi) M_1^2 - \xi M_2^2$. The important fact
here is that, even after integration on the transverse momenta of all particles,
still a T-odd PFF survives, $H_1^{\open}$, because there is another transverse
momentum, ${\bf R}_{_T}$, available to generate an azimuthal asymmetry. Luckily,
$H_1^{\open}$ is also chiral-odd and, therefore, it represents a perfect partner
to isolate transversity. In fact, the corresponding SSA is~\cite{noi3}
\begin{eqnarray}
A_{_{OT}}^{\sin \phi} &= &\frac{\int d\phi_{_S} d\phi_{_R} d\xi \sin
(\phi_{_S}+\phi_{_R}) [d\sigma (p^\uparrow) - d\sigma (p^\downarrow)]}{\int
d\phi_{_S} d\phi_{_R}d\xi [d\sigma (p^\uparrow) + d\sigma (p^\downarrow)]}
\nonumber \\
&\propto &|{\bf S}_{_T}|\, \frac{\sum_f \, e_f^2 h_1^f(x) \int d\xi d\phi_{_R}
\, \frac{|{\bf R}_{_T}|}{2M_h}\, H_1^{\open f}(z,\xi,M_h^2)}{\sum_f\,e_f^2
f_1^f(x) D_1^f(z,M_h^2)} \nonumber \\
&\equiv &|{\bf S}_{_T}|\,\frac{\sum_f\,e_f^2 h_1^f(x)H_{1(R)}^{\open
f}(z,M_h^2)}{\sum_f\,e_f^2 f_1^f(x) D_1^f(z,M_h^2)} \; , 
\label{eq:2hssa}
\end{eqnarray}
where the specific moment $H_{1(R)}^{\open f}$ is involved. The very same moment
appears when considering the correspoding $e^+e^- \rightarrow (\pi \pi)_{jet1}
(\pi \pi)_{jet2} X$ process at leading twist and looking for an azimuthal
asymmetry in the position of the pion pair planes with respect to the lab
frame~\cite{noi4}. Even if a factorization theorem is not yet proven for IFF, the
universality holds, at least at leading twist, allowing to extract informations
on the unknown function $H_1^{\open}$. Moreover, because of the disappearance of
any transverse momenta of involved particles all the difficulties arising for
the Collins effect are here overcome: collinear factorization holds, avoiding
any dilution of SSA by Sudakov form factors; the analogue of the Sivers effect
is here absent, keeping formulae simpler; in the corresponding annihilation
$pp^\uparrow \rightarrow (\pi \pi) X$ the elementary cross section does not
depend on transverse parton momenta, so that the result is truly a leading-twist
one~\cite{noi-upcoming}.

Another advantage of IFF is the possibility of studying in some detail the FSI
responsible for the T-odd structures. For example, if the two unpolarized
hadrons are two pions, in their center-of-mass frame the IFF can be expanded in
relative partial waves retaining the main contributions, which for two pions are
the $s$ and $p$ waves~\cite{noi5}. Therefore, since FSI arise from the interference
of different channels with different phases, two possible sources of
interference appear: the $s-p$~\cite{jaffe-iff} and the $p-p$ ones. Both these 
components of $H_1^{\open}$ act in the SSA of eq.~\ref{eq:2hssa} and can be 
disentangled by a suitable selection of the integration phase space~\cite{noi5}. In 
particular, the latter is formally and closely related to the fragmentation of 
spin-1 objects, like the $\rho$ polarized fragmentation functions~\cite{rho}.

When considering two leading unpolarized hadrons inside the jet of the current
fragmentation region, another interesting azimuthal asymmetry can be built for
the $e^+e^-\rightarrow (\pi \pi)_{jet1} (\pi\pi)_{jet2} X$ process, which
involves the helicity IFF $G_1^\perp$~\cite{noi4}:
\begin{eqnarray}
\langle \cos 2(\phi_{_R} - \bar{\phi}_{_R})\rangle &= &\frac{\int d\xi
d\phi_{_R} d\bar{\xi} d\bar{\phi}_{_R}d{\bf q}_{_T}\, \cos 2(\phi_{_R} -
\bar{\phi}_{_R}) d\sigma}{\int d\xi d\phi_{_R} d\bar{\xi} d\bar{\phi}_{_R} d{\bf
q}_{_T}\, d\sigma} \nonumber \\
&\propto &\frac{\sum_f\,e_f^2 \int d\xi d\phi_{_R} d{\bf k}_{_T}\, {\bf
k}_{_T}\cdot {\bf R}_{_T}\, G_1^{\perp f} (z,\xi,{\bf k}_{_T}^2,{\bf R}_{_T}^2,
{\bf k}_{_T}\cdot {\bf R}_{_T})}{\sum_f\,e_f^2 \int d\xi d\phi_{_R} 
d{\bf k}_{_T}\, D_1^f (z,\xi,{\bf k}_{_T}^2,{\bf R}_{_T}^2,{\bf k}_{_T}\cdot 
{\bf R}_{_T})} \nonumber \\
& &\quad \times \frac{\int d\bar{\xi} d\bar{\phi}_{_R} d{\bf \bar{k}}_{_T}\, 
{\bf \bar{k}}_{_T}\cdot {\bf \bar{R}}_{_T}\, \bar{G}_1^{\perp f} (\bar{z},
\bar{\xi}, {\bf \bar{k}}_{_T}^2,{\bf \bar{R}}_{_T}^2,{\bf \bar{k}}_{_T}\cdot 
{\bf \bar{R}}_{_T})}{\int d\bar{\xi} d\bar{\phi}_{_R} d{\bf \bar{k}}_{_T}\, 
\bar{D}_1^f (\bar{z},\bar{\xi},{\bf \bar{k}}_{_T}^2,{\bf \bar{R}}_{_T}^2,
{\bf \bar{k}}_{_T}\cdot {\bf \bar{R}}_{_T})} \nonumber \\
&\equiv &\frac{\sum_f\,e_f^2 G_{1\otimes}^{\perp
f}(z,M_h^2)\,\bar{G}_{1\otimes}^{\perp f}(\bar{z},\bar{M}_h^2)}{\sum_f \, e_f^2
D_1^f(z,M_h^2)\, \bar{D}_1^f(\bar{z},\bar{M}_h^2)} \; .
\label{eq:e+e-ssa}
\end{eqnarray}
When integrating directly on ${\bf k}_{_T}$, $G_1^\perp$ vanishes because of
parity invariance. However, the asymmetry in the azimuthal relative position of
the planes of the two pion pairs involves the specific nonvanishing "moment"
$G_{1\otimes}^\perp$, which allows for a relation between the longitudinal
polarization of the fragmenting quark and the transverse relative motion of the
hadron pair. This is a unique feature of IFF describing fragmentation into two
unpolarized hadrons. The helicity analyzer $G_1^\perp$ occurs weighted by the
$({\bf k}_{_T} \times {\bf R}_{_T})$ factor in the leading-twist projection;
therefore, it is the closest analogue to the definition of longitudinal jet
handedness~\cite{handedness}. It could be extracted from the cross section of the 
$e\vec p \rightarrow e'(\pi \pi)X$ process, where it appears at leading twist 
convoluted with the known helicity distribution $g_1$. Assuming universality, it 
could be plugged inside eq.~\ref{eq:e+e-ssa}: a nonvanishing azimuthal asymmetry 
could then be the signal that a violation is taking place in the CP symmetry of the
two back-to-back jets, maybe due to the effect of the nonperturbative vacuum of
QCD~\cite{kharzeev}.

\section{Transversity from Drell-Yan processes}
\label{sec:dy}
\begin{figure}[h]
 \vspace{4.0cm}
\includegraphics{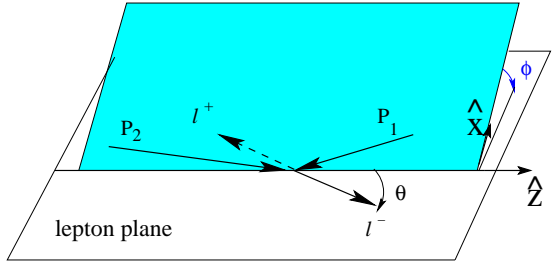}	 
\caption{\it The Collins-Soper frame for Drell-Yan processes.
\label{fig:dyframe} }
\end{figure}

As already anticipated in sec.~\ref{sec:Collins}, the polarized Drell-Yan process
$p^\uparrow p^\uparrow \rightarrow l^+ l^- X$ was the first suggested in order to
extract the transversity at leading order~\cite{poldy}. In the socalled 
Collins-Soper frame, the $\hat{z}$ axis is obtained by the "average" direction of 
the two annihilating hadron momenta, which lie in the hadronic plane rotated around
$\hat{z}$ by an azimuthal angle $\phi$ with respect to the lepton plane where the
two leptons are emitted back-to-back at an angle $\theta$ with respect to $\hat{z}$
(see fig.~\ref{fig:dyframe}). 

\begin{figure}[t]
 \vspace{4.5cm}
\includegraphics{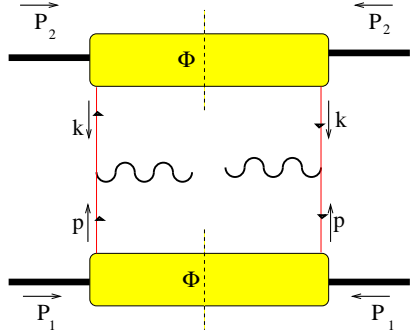}	 
\caption{\it The leading contribution to the Drell-Yan process.
\label{fig:dylead} }
\end{figure}

If the dominant contribution comes from the diagram depicted in
fig.~\ref{fig:dylead}, then the following DSA, 
\begin{eqnarray}
A_{TT} &= &\langle \cos (\phi_{_{S_1}}+\phi_{_{S_2}})\rangle = 
\frac{\int d{\bf q}_{_T} \, \cos (\phi_{_{S_1}}+\phi_{_{S_2}})\, [ d\sigma
(p^\uparrow p^\uparrow ) - d\sigma (p^\uparrow p^\downarrow )]}
{\int d{\bf q}_{_T} \, [ d\sigma (p^\uparrow p^\uparrow ) + d\sigma 
(p^\uparrow p^\downarrow )]} \nonumber \\
&= &|{\bf S}_{_{T_1}}||{\bf S}_{_{T_2}}|\, \frac{\sin^2 \theta \, \cos
2\phi}{1+\cos^2 \theta}\,
\frac{\sum_f\,e_f^2\,h_1^f(x_1)\,\bar{h}_1^f(x_2)}
{\sum_f\,e_f^2\,f_1^f(x_1)\,\bar{f}_1^f(x_2)}\; ,
\label{eq:dsaTT}
\end{eqnarray}
shows a factorized combination of transversities for the annihilating quark and
antiquark at leading twist. The angles $\phi_{_{S_1}}, \phi_{_{S_2}},$ define the
azimuthal position of the transverse polarization vectors of the two annihilating
hadrons with respect to the lepton plane, while $x_{1/2} = Q^2 / (2P_{1/2}\cdot q)$
and ${\bf q}_{_T}$ lies in the $(\hat{x}, \hat{z})$ plane of
fig.~\ref{fig:dyframe}. Unfortunately, in a NLO simulation $A_{TT}$ turns out to be
suppressed by the Soffer inequality and QCD evolution~\cite{aTT,drago}, and, 
moreover, it involves the transversity of an antiquark in a transversely polarized 
proton. The latter difficulty could be overcome at future GSI-HESR with polarized 
antiproton beams, where the asymmetry would involve transversities of valence 
partons anyway~\cite{goeke-GSI,drago-GSI}. 

Also the single polarized and unpolarized Drell-Yan cross sections show interesting
combinations at leading twist. For the $pp^\uparrow \rightarrow l^+ l^- X$ process,
the leading-twist cross section displays, among others, the following
contributions~\cite{boer}:
\begin{eqnarray}
\frac{d\sigma}{dx_1 dx_2 d\Omega d{\bf q}_{_T}} &\propto &|{\bf S}_{_{T_1}}|\, 
\sum_f\, \Bigg\{ \sin (\phi - \phi_{_{S_1}})\, {\cal F} \left[ \hat{\bf h}\cdot
{\bf p}_{_{T_1}}\, f_{_{1T}}^{\perp f}(x_1,{\bf p}_{_{T_1}})\, \bar{f}_1^f
(x_2,{\bf p}_{_{T_2}})\right] \nonumber \\
& &+ ... \sin (\phi + \phi_{_{S_1}})\, {\cal F}\left[ \hat{\bf h}\cdot {\bf
p}_{_{T_2}}\, h_1^f (x_1,{\bf p}_{_{T_1}})\, \bar{h}_1^{\perp f}(x_2,{\bf
p}_{_{T_2}}) ... \right] \Bigg\} \; . \nonumber \\
& &{} \label{eq:ssady}
\end{eqnarray}
The first term can be isolated by an azimuthal asymmetry similar to the one of the
Sivers effect, in order to extract $f_{_{1T}}^\perp$ and compare it with the result
of semi-inclusive processes (eq.~\ref{eq:ssa}): the different behaviour of the
gauge link operator in SIDIS and Drell-Yan processes leads to the interesting
conjecture that a possible change of sign could take place, posing a question about
the assumed universality of T-odd PDF~\cite{collins2,link-amst} (see also
ref.~\cite{metz}).

The second term in eq.~\ref{eq:ssady} contains the transversity and can be isolated
by an asymmetry similar to the Collins effect. However, $h_1$ appears convoluted
with a second unknown function, $h_1^\perp$, which also contributes to the SSA for
the $pp^\uparrow \rightarrow \pi X$ process. Interestingly, this function
contributes also to the unpolarized Drell-Yan cross section at leading
twist~\cite{boer}, i.e.
\begin{eqnarray}
\frac{d\sigma}{dx_1 dx_2 d\Omega d{\bf q}_{_T}} &\propto &\sum_f\, \Bigg\{ {\cal
F}\left[ f_1^f \, \bar{f}_1^f\right] ... \nonumber \\
& &\quad + \cos 2\phi \, {\cal F}\left[ A({\bf p}_{_{T_1}}, {\bf p}_{_{T_2}})\,
h_1^{\perp f}(x_1,{\bf p}_{_{T_1}})\, \bar{h}_1^{\perp f}(x_2,{\bf p}_{_{T_2}})\right] 
... \Bigg\} \; , \nonumber \\
& &{} \label{eq:unpoldy}
\end{eqnarray}
where $A({\bf p}_{_{T_1}}, {\bf p}_{_{T_2}})$ is some function of the transverse
momenta of the annihilating partons. The crucial remark is that $h_1^\perp$ can
naturally explain the observed sizeable azimuthal asymmetry in unpolarized
Drell-Yan cross sections. The most general parametrization of such cross sections
at leading order looks like 
\begin{equation}
\frac{1}{\sigma}\,\frac{d\sigma}{d\Omega}\,\propto\,1+\lambda\,\cos^2\theta +
\mu\,\sin^2\theta \cos \phi + \frac{\nu}{2}\,\sin^2\theta\cos 2\phi + o(\alpha_s)
\; ,
\label{eq:unpolxsect}
\end{equation}
where the data suggest $\lambda \sim 1$ and $\mu \ll \nu \sim 30$\%~\cite{na10}, 
while the perturbative QCD gives $\lambda \sim 1, \mu \sim \nu \sim 0$. Neither 
higher twists, nor factorization-breaking terms of NLO contributions are able to 
justify such a big azimuthal asymmetry~\cite{varie}, while eq.~\ref{eq:unpoldy} 
easily accounts for it because it contains a leading twist term $\sim \cos 2\phi$ 
and no contributions $\sim \cos \phi$. A simple explanation for this feature is the 
fact that, owing to its probabilistic interpretation (see 
fig.~\ref{fig:f1Tperph1perp}), each $h_1^\perp$ carries one unit of quark orbital 
angular momentum $L_z$, introducing then a cosinusoidal dependence on twice the 
azimuthal angle $\phi$.

In conclusion, the unpolarized Drell-Yan process allows for the extraction of a
chiral-odd T-odd PDF, $h_1^\perp$, that can help in disentangling the transversity
$h_1$ in the corresponding single-polarized Drell-Yan process and, in turn, to
compare it with the one that can be extracted in SIDIS or annihilation processes.
Even unpolarized observables can contribute to the study of spin structure of the
nucleon. If beams of protons and antiprotons are used, as it will be at GSI-HESR,
the Drell-Yan azimuthal asymmetries will not be suppressed by nonvalence-like
contributions. 

\begin{figure}[t]
 \vspace{7.0cm}
\includegraphics{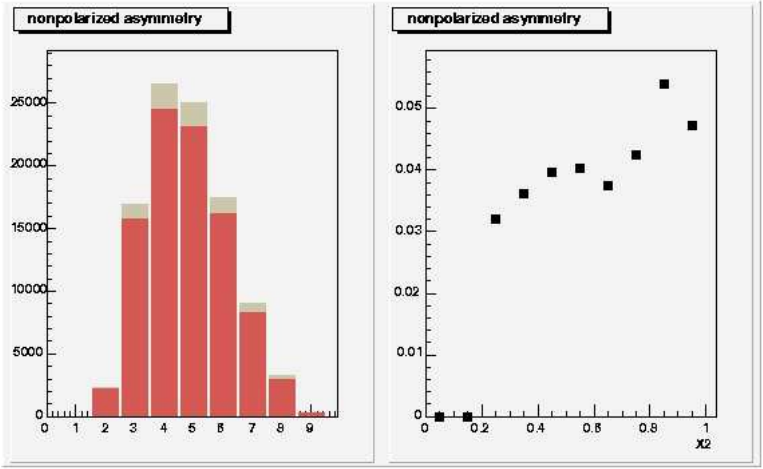}	 
\caption{\it Left panel: cross section for unpolarized Drell-Yan with antiproton
beams in the kinematics of the ASSIA proposal; different histogram colors for
different signs in the $\cos 2\phi$ azimuthal term. Right panel: the corresponding
azimuthal asymmetry.
\label{fig:mcout} }
\end{figure}

\section{Monte-Carlo simulations}
\label{sec:mc}
In this section, some Monte-Carlo simulations are presented for the unpolarized and
single-polarized Drell-Yan azimuthal asymmetries, involving proton and antiproton
beams with the kinematical conditions discussed in the ASSIA 
proposal~\cite{andrea}. Very briefly, the considered process is 
$\bar{p} p^{(\uparrow )}\rightarrow \mu^+ \mu^- X$ with an antiproton beam of 
$E_{\bar{p}}= 40$ GeV. The proton target is obtained by a $NH_3$ target with a 
dilution factor of ${\textstyle \frac{1}{4}}$. When the target is polarized, a 
further 85\% reduction is applied. The center-of-mass energy is 
$s \sim 2M_p E_{\bar{p}} \sim (9{\rm GeV})^2$ and $\tau = x_1 x_2 = Q^2/s \leq 1$ 
for the invariant mass $Q \leq 9$ GeV. Finally, we have $-0.7 \leq x_{_F} = x_1 - 
x_2 \leq 0.7$ for the invariant fraction of total longitudinal momentum with
respect to the maximum available longitudinal momentum for the system of 
annihilating partons.

The simulation consists of 480.000 events distributed according to the cross
section
\begin{eqnarray}
\frac{1}{\sigma}\,\frac{d\sigma}{d\Omega dx_1 dx_2 d\phi_{_{S_2}}} &= &1 +
\cos^2\theta + \frac{\nu (x_1,x_2,p_{_T})}{2}\,\sin^2\theta\,\cos 2\phi \nonumber
\\
& &\quad +|{\bf S}_{_{T_2}}|\, \sin^2\theta \, \sin (\phi + \phi_{_{S_2}})\,
A(x_1,x_2,p_{_T}) \; , 
\label{eq:mcxsect}
\end{eqnarray}
all the variables are represented in fig.~\ref{fig:dyframe} and $p_{_T}$ is the
transverse momentum of the lepton pair. The latter is cut below 1 GeV$/c$ as well
as the invariant mass of the muon pair, namely $M_{\mu\mu} > 4$ GeV. 

In fig.~\ref{fig:mcout}, the left panel shows the cross section of
eq.~\ref{eq:mcxsect} for $|{\bf S}_{_{T_2}}|=0$ and binned in $x_2$ with positive
and negative $\cos 2\phi$. The corresponding azimuthal asymmetry is shown in the
right panel, which is proportional to $\nu (x_1, x_2, p_{_T})$ and is parametrized
according to ref.~\cite{conway}. The obtained asymmetry around 5\% seems 
measurable. 

\begin{figure}[t]
 \vspace{7.0cm}
\includegraphics{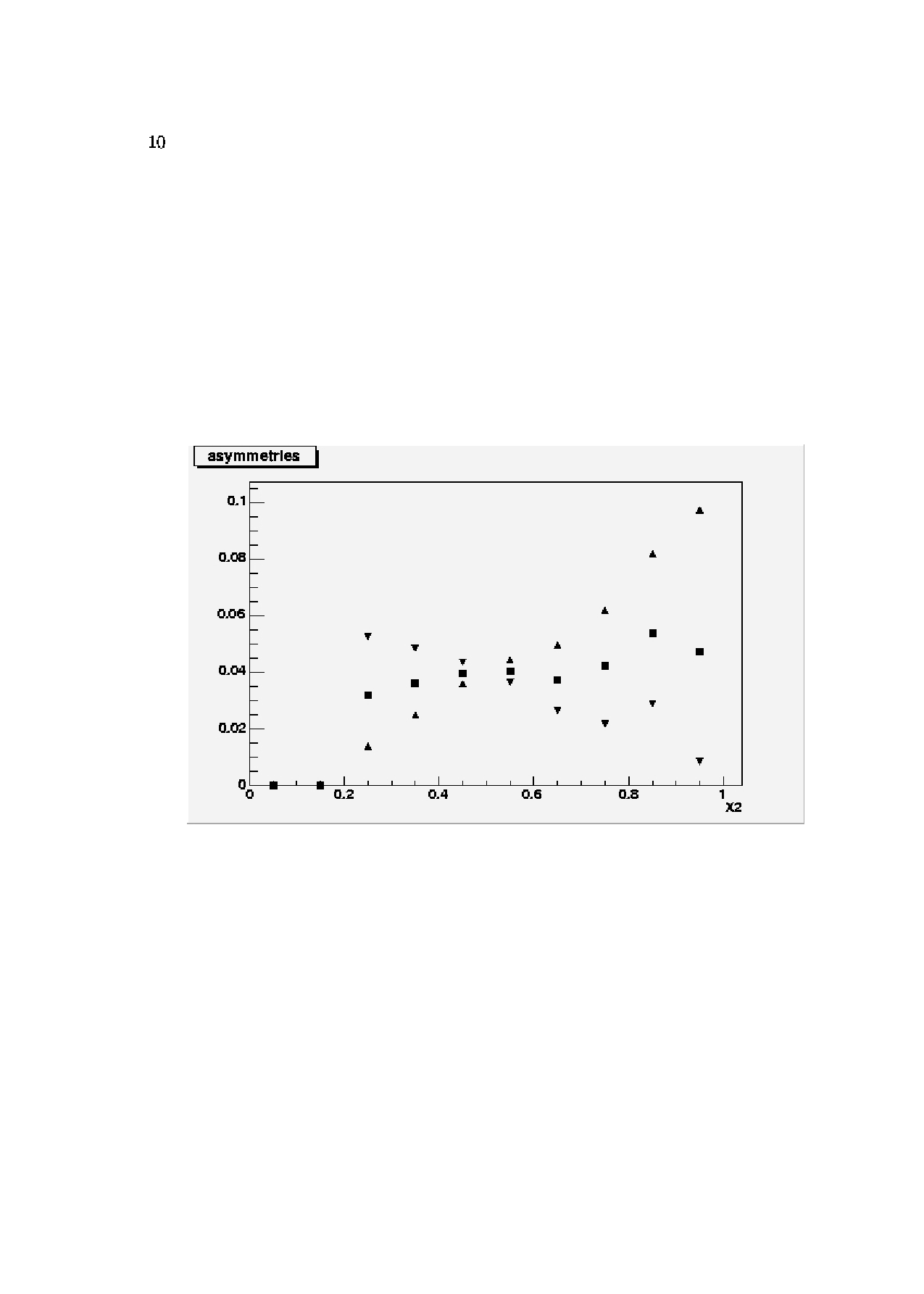}	 
\caption{\it The SSA for polarized Drell-Yan with antiproton beams in the
kinematics of the ASSIA proposal. Upward triangles, squares, and downward triangles
for different choices of the $x_2$ dependence (see text).
\label{fig:mcout1} }
\end{figure}

Next, in fig.~\ref{fig:mcout1} the same simulation in the same conditions is
applied to the cross section of eq.~\ref{eq:mcxsect} with $|{\bf S}_{_{T_2}}|\neq
0$ and with different factorized behaviours for $A(x_1,x_2,p_{_T})$: upward
triangles for $2(1-x_2)\,A(x_1,p_{_T})$, squares for $1\cdot A(x_1,p_{_T})$,
downward triangles for $2x_2\,A(x_1,p_{_T})$. The function $A(x_1,p_{_T})$ is
cancelled in the ratio of cross sections defining the azimuthal asymmetry;
therefore, it needs not to be specified. The $x_2$ dependences represent three
different "wild guesses" for the factorized ratio $h_1(x_2)/f_1(x_2)$ that appears
in the asymmetry at leading twist. Despite the adopted crude approximations, the
azimuthal asymmetry is sensitive enough to distinguish between these three choices,
giving at the same time an average measurable result of around 5\%. 

\vspace{1cm}

In conclusion, Monte-Carlo simulations running antiproton beams on (polarized)
proton targets at the kinematics of the ASSIA proposal at the
GSI-HESR facility, seem to produce a reasonable number of Drell-Yan events, such
that an average 5\% azimuthal asymmetry is observed using realistic cut-offs and a
sample of almost half million of events.

\section{Ackonwledgements}
I thank A.~Bianconi for making available his calculations of the Monte-Carlo
simulations reported in sec.~\ref{sec:mc}.

%
%
%

%
\end{document}